\documentstyle[amssymb,aps]{revtex}

\begin{document}
\draft

\twocolumn[\hsize\textwidth\columnwidth\hsize\csname @twocolumnfalse\endcsname
\title{Ultracold atoms interacting
 with a sinusoidal mode of a  high {\it Q} cavity}
\author{J.~C.~Retamal${}^\dagger$, E.~Solano${}^\dagger$$^\dagger$$^*$ and
 N.~Zagury${}^*$}
\address{$^\dagger$ Departamento de F\'\i sica, Universidad de Santiago de Chile,
 Casilla 307, Correo 2, Santiago, Chile}
\address{$^\dagger$$^\dagger$Seccion F\'\i sica, Departamento de Ciencias,
 Pontificia Universidad Catolica del Peru, Ap. 1761, Lima, Peru}
\address{$^*$Instituto de F\'\i sica, Universidade Federal do Rio de
Janeiro,
Caixa Postal 68528, 21945-970 Rio de Janeiro, RJ, Brazil}
\date{\today }
\maketitle
\begin{abstract}
We consider the interaction of two level ultracold atoms resonant with 
a sinusoidal mode of the electromagnetic field in a high $Q$ cavity. We found that 
well resolved resonances appear in the transmission coefficients 
even for actual interaction and cavity parameters. The probability of emission of one photon and the probability of transmission  of an atom, when  number or coherent states
are initially  present in the cavity, are  discussed. The interplay between the increasing width of the  resonances and multi-peak steady-state photon-statistics are also studied. Furthermore, we compare our results with those of a  constant field mode. 
\end{abstract}
\pacs{PACS numbers: 42.50.-p, 32.80.-t, 42.50.Ct, 42.50.Dv }

\vskip2pc]

\section{Introduction}

Recent developments in cooling of neutral atoms\cite{cooling} have called
the attention on the interaction of ultracold atoms with microwave cavities 
\cite{englert,scully,haroche,scully1,scully2,scully3}. Very cold atoms
(kinetic energies much smaller than the atom-field interaction energy ) have
a peculiar behavior when interacting with a cavity, they are strongly
reflected unless the ratio of the size of the cavity and the atomic wave
length associated to the interaction energy are close to certain resonant
values. For example, it has been shown \cite{scully,scully1} that ultracold
atoms of mass $m$ may be totally reflected by a cavity of size $L$
containing $n$ photons, in a constant field mode, unless the Rabi frequency $%
\Omega _{n}$ is close to $2\hbar (j\pi /L)^{2}/(2m),$ $j=0,1,2...$, when
only approximately half of the atoms are reflected.

Models for the interaction of atoms with a microwave cavity usually do not
take into account the spatial dependence of the interaction coming from the
electric field profile\cite{Meystre}. This is justified because, in general,
we are interested in the atom-field state {\it after} the atom has crossed
the cavity and are dealing with atomic {\it thermal velocities}, which are
exceedingly high compared with the recoil ones experienced by the atom
inside the cavity.

In order that we may observe any effect on the center-of-mass motion of the
atoms, their velocities have to be reduced to very low values in such a way
that even the microscopic inversion of momentum of the center-of-mass could
take place. Actually, the only way for an atom to experience mechanical
action on its center-of-mass is through the mechanism of exchange of
momentum with the whole cavity. At high atomic velocities the energy
exchange between the field and the atoms takes place through the Rabi
flopping of the internal atomic levels. At very low atomic velocities the
spatial variation of the electromagnetic field acting on the atoms provides
an additional driving mechanism which lies in the momentum exchange of the
atom with the cavity.

The aim of the present work is to investigate the effect of a sinusoidal
electric field mode when ultracold atoms are sent into a cylindrical
microwave cavity for actual experimental parameters. We will show that under
these conditions the behavior of the relevant physical quantities may be
quite different from those of a discontinuous field mode, although still
preserving resonant features if the atoms are cold enough.

\begin{figure}[tbp]
\end{figure}

\section{The model}

Let us assume that a two level atom is travelling along the $z$-direction in
the way of a microwave high $Q$ cavity of length $L$, containing a quantized
field mode in resonance with the two level atomic transition $|e\rangle
\Leftrightarrow |f\rangle $.

The Hamiltonian describing the motion of the two level atom in the $z$
-direction is given, in the interaction representation, by 
\begin{equation}
H=\frac{p^{2}}{2m}+\frac{\hbar ^{2}\kappa ^{2}}{2m}u(z)(a\sigma ^{\dagger
}+a^{\dagger }\sigma ).  \label{Hamiltonian}
\end{equation}
Here $\sigma \equiv |f\rangle \langle e|,$ $a$ ($a^{\dagger }$) is the
annihilation (creation) operator of the electromagnetic field mode whose
spatial modulation pattern is $u(z)$ and $\hbar \kappa ^{2}/(2m)$ is half
the Rabi frequency associated with the transition $|f\rangle \leftrightarrow
|e\rangle $ in the vacuum. $u(z)$ is normalized in such a way that the area
under the mode is $L$. In refs. \cite{scully,scully1} $u(z)$ is taken as a
square function of height $1$ and width $L$ (mesa function). In this case it
can be shown that the transmission of ultracold atoms through a cavity
containing $n$ photons is very small unless 
\begin{equation}
\kappa _{n}\equiv \kappa \sqrt[4]{n+1}
\end{equation}
is a multiple of $\pi /L.$ In ref. \cite{scully2} the $\text{sech }^{2}(z/L)$
and sinusoidal modes were considered and the authors have concluded that,
for high values of $L,$ the resonances referred above are smeared out. Here
we will discuss in more detail the sinusoidal mode having just one antinode.
We will show that in this case, resonances still exist for high values of $%
\kappa _{n}L$ as long as the atoms are cold enough.

The Hamiltonian given in Eq. \ref{Hamiltonian} can be conveniently written
in the dressed state basis as:

\begin{equation}
H=\sum [\frac{p^{2}}{2m}\pm V_{n}(z)]|\pm ,n\rangle \langle \pm ,n|,
\label{Hamiltonian2}
\end{equation}
where $V_{\pm n}(z)$ is given by 
\begin{equation}
V_{\pm n}(z)=\pm \frac{\hbar ^{2}\kappa _{n}^{2}}{2m}u(z),
\end{equation}
and

\begin{equation}
|\pm ,n\rangle =(|e,n\rangle \pm |f,n+1\rangle )/\sqrt{2}.
\end{equation}

In the dressed state basis our problem is equivalent to that of a particle
being scattered by the potentials $V_{\pm n}(z)$. The cavity acts as a
barrier when the atom-field state is $|+n\rangle $ and acts as a well when
the atom-field state is $|-n\rangle $.

Before the atom reaches the cavity containing the field $\sum c(n)|n\rangle$%
, the initial atomic state is given by the product of a wave packet in
momentum space $\langle z|\phi \rangle $=$\int dk\phi (k)e^{ikz}$ times the
internal energy state of the atom that, for definitiveness, we assume to be
the upper state $|e\rangle $. After the atom interacts with the cavity the
total atom-cavity state is (for $|z|>L/2$):

\begin{eqnarray}
&&\sum c(n)\int dk\phi (k)\text{exp}(-i\frac{\hbar ^{2}k^{2}}{2m} \tau
)[(e^{ikz}t_{+n}(k)+ e^{-ikz}r_{+n}(k))  \nonumber \\
&&|+n\rangle +(e^{ikz}t_{-n}(k)+e^{-ikz}r_{-n}(k))|-n\rangle ]\text{,}
\end{eqnarray}
where the reflection, $r_{\pm n}$, and transmission, $t_{\pm n}$,
coefficients are associated with the scattering eigenfunctions $\varphi
_{k}^{\pm }(z)$. These functions are solutions of the time independent
Schr\"{o}dinger equation : 
\begin{equation}
\left[ \frac{d^{2}}{dz^{2}}+k^{2}\mp \kappa _{n}^{2}u(z)\right] \varphi
_{k}^{\pm }(z)=0,  \label{Schr}
\end{equation}
where $\hbar k$ is the momentum of the incident particle.

In this paper we are interested in discussing a situation where the
longitudinal $z$-dependence of the electric field is sinusoidal:

\begin{equation}
u(z)=\theta (L/2-|z|)(\pi /2)\cos (\frac{\pi z}{L}).
\end{equation}
This is the case of a cylindrical cavity tuned either to a TE$_{mn1}$ or a
TM $_{mn1}$ mode\cite{Jackson}.

We may try to obtain the values of $r_{\pm n}(k)$ and $t_{\pm n}(k)$ by
numerically integrating a complex solution of Eq. \ref{Schr} for runaway
asymptotic conditions. An alternative way, where we found numerical and
analytical advantages, is to consider instead the even $(\varphi _{e}^{\pm
n}(z))$ and odd $(\varphi _{o}^{\pm n}(z))$ real (up to an overall complex
constant) eigensolutions of Eq. \ref{Schr}, such that for $z<-L/2$\cite
{Eberly}

\begin{equation}
\varphi _{e,o}^{\pm n}(z)\propto \text{cos}(kz-\delta _{e,o}^{\pm n}).
\label{phaseshifts}
\end{equation}
Here $\delta _{e}^{\pm n}(k)$ $(\delta _{o}^{\pm n}(k))$ is the phase shift
associated to the even (odd) eigensolution of the Schr\"{o}dinger equation.
In terms of these phase shifts the reflection and transmission coefficients
may be written as

\begin{eqnarray}
r_{\pm n}(k) &=&\frac{e^{2i\delta _{e}^{\pm n}}+e^{^{2i\delta _{o}^{\pm n}}} 
}{2}  \nonumber \\
&&  \label{trans} \\
t_{\pm n}(k) &=&\frac{e^{2i\delta _{e}^{\pm n}}-e^{2i\delta _{o}^{\pm n}}}{2}
.  \nonumber
\end{eqnarray}

The phase shifts are related to the logarithmic derivatives at $z=-L/2$ by 
\begin{equation}
\beta _{e,o}^{\pm n}=k\text{ tan}{(\kappa }_{n}{L/2+}\delta _{e,o}^{\pm n}).
\label{beta}
\end{equation}
Notice that $\beta _{e,o}^{\pm n}$ are real in contrast with the complex
logarithmic derivative of a typical, right or left, propagating solution.
>From Eqs. \ref{trans} and \ref{beta} it is easy to show that: 
\begin{equation}
r_{\pm n}(k)=\frac{k^{2}+\beta _{e}^{\pm n}\beta _{o}^{\pm n}}{(k-i\beta
_{e}^{\pm n})(k-i\beta _{o}^{\pm n})}e^{-ikL}  \label{reflec}
\end{equation}
and 
\begin{equation}
t_{\pm n}(k)=\frac{ik(\beta _{e}^{\pm n}-\beta _{o}^{\pm n})}{(k-i\beta
_{e}^{\pm n})(k-i\beta _{o}^{\pm n})}e^{-ikL}.  \label{trans1}
\end{equation}

In the case of the potential barrier, for $\kappa _{n}L\gg 1$ and $k\ \ll
\kappa _{n},$ the transmission coefficient $t_{+n}$ is very small and close
to zero. In the case of the potential well, and for $k\ \ll \kappa _{n}$ the
transmission coefficient $t_{-n}$ is close to zero unless $\beta
_{e}^{-n}\beta _{o}^{-n}=-k^{2}$, that is for very small $\beta _{e}^{-n}$or
very small $\beta _{o}^{-n}$. For a fixed value of $k/\kappa _{n}$, these
resonances in the transmission coefficient occur at special values of $%
\kappa _{n}L$, which would correspond to the appearance of a bound state
with $k^{2}=0$ ( notice that the transmission coefficient has a pole at $%
k=i\beta _{e,o}^{-n}$).

Assume that initially there are $n$ photons inside the cavity and the atom
is incident from the left in the upper state $|e\rangle $. After the atom
interacts with the cavity, it can be transmitted in the upper state,
transmitted in the lower state, reflected in the upper state or reflected in
the lower state. The probabilities that these events occur are denoted by $%
T_{e}^{n}$, $T_{f}^{n}$, $R_{e}^{n}$and $R_{f}^{n}$, respectively and given
by:

\begin{equation}
T_{e}^{n}=\frac{1}{4}\left| t_{+n}+t_{-n}\right| ^{2},\quad T_{f}^{n}=\frac{%
1 }{4}\left| t_{+n}-t_{-n}\right| ^{2}
\end{equation}

and

\begin{equation}
R_{e}^{n}=\frac{1}{4}\left| r_{+n}+r_{-n}\right| ^{2},\quad R_{f}^{n}=\frac{%
1 }{4}\left| r_{+n}-r_{-n}\right| ^{2}.
\end{equation}

Of course $T_{e}^{n}+T_{f}^{n}+R_{e}^{n}+R_{f}^{n}=1$. We will show in the
next section that for $k/\kappa _{n}\ \ll 1$, $t_{+n}\cong 0$. In this case $%
T_{e}^{n}\approx T_{f}^{n}\approx |t_{-n}|^{2}/4$. Due to this last
approximate relation and the unitarity condition we choose to discuss below
only two quantities of physical interest:\ the probability of emission of a
photon, $P_{\text{em }}^{n}=T_{f}^{n}+R_{f}^{n}$, and the probability of
transmission of an atom, $T^{n}=T_{e}^{n}+T_{f}^{n}$. We notice also that
when $r_{-n}\cong 0$, $P_{\text{em}}^{n}\cong T^{n}$.

If we initially have a photon number distribution given by $|c_{n}|^{2}$,
the total probability of photon emission is $P_{\text{em }}=\sum
|c_{n}|^{2}P_{\text{em }}^{n}$and the total probability of atomic
transmission is $T=\sum |c_{n}|^{2}T_{\text{ }}^{n}$. These quantities
should be easily measurable by detecting the internal energy levels of the
reflected and transmitted atoms. As we will show, these quantities depend
strongly on the shape of the field inside the cavity, as long as $k/\kappa
_{n}\ \ll 1$.

\section{Analytical Results}

The constant field mode ($u(z)=1$, for $|z|<L/2)$ have been discussed in
refs. \cite{englert,scully,scully1}. For $k/\kappa _{n}\ \ll 1$ we have $%
\beta _{e}^{+n}\approx -\kappa _{n}$ tanh$(\kappa _{n}L/2)$, $\beta
_{o}^{+n}\approx -\kappa _{n}$ coth$(\kappa _{n}L/2)$, $\beta
_{e}^{-n}\approx \kappa _{n}$ tan$(\kappa _{n}L/2)$ and $\beta
_{o}^{-n}\approx \kappa _{n}$ cot$(\kappa _{n}L/2)$, so that the resonances
appear when $\kappa _{n}L$ equals to an integer multiple of $\pi $\cite
{englert,scully,scully1}. In this case the resonances are very sharp, their
width being approximately constant and equal to $4k/\kappa _{n}$ even for
large $\kappa _{n}L$.

In the case of the sinusoidal mode, as long as $\kappa _{n}L$ is small, it
is easy to obtain the even and odd eigensolutions of Eq. \ref{Schr} by
numerical integration . This can be done easily and with confidence for $%
\kappa _{n}L$$\lesssim 100\pi $. If we wish to predict values closer to real
experimental situations we need to consider higher values of $\kappa _{n}L$,
which are typically of the order of $O(10^{5}-10^{6})$ for Rydberg atoms. In
these cases the numerical solutions do not converge rapidly for the $%
|-n\rangle $ channel and we found easier, and more instructive, to use a
WKB-like approximation for calculating the $\delta _{e}^{\pm n}$ and $\delta
_{o}^{\pm n}$ phase shifts or, equivalently, the logarithmic derivatives $%
\beta _{e,o}^{\pm n}$.

In the case of the potential barrier, and for $k/\kappa _{n}\ \ll 1$, we may
use the WKB$\;$approximation\cite{Morse} for calculating $\beta _{o,e}^{+n}$
. We get

\begin{equation}
\beta _{e,o}^{+n}\cong -k(1\mp \Theta ),  \label{beta+}
\end{equation}
where the $-$ and $+$ signs inside the parenthesis refer to the even and odd
solution and 
\begin{equation}
\Theta =\text{exp}(-\int_{-a}^{a}\sqrt{-k^{2}+\kappa _{n}^{2}u(z)}dz),
\label{Theta}
\end{equation}
where $\pm a$ are the turning points. Substituting the values for $\beta
_{o,e}^{+n}$ in Eqs. \ref{reflec} and \ref{trans1} we get $r_{+n}(k)$$\simeq 
$$-ie^{-ikL}$ and $t_{+n}(k)$$\simeq i\Theta e^{-ikL}$. For $k/\kappa _{n}\
\ll 1,$ $a\simeq L/2$ and $\Theta \simeq $exp$(-4.19\kappa _{n}L),$ which is
very small for $\kappa _{n}L>1.$

Inside the cavity the solutions $\varphi _{e,o}^{-n}(z)$ for the potential
well may be written, {\it whenever the semiclassical approximation holds},
as 
\begin{eqnarray}
\varphi _{e}^{-n}(z) &\approx &\frac{1}{\sqrt{q(z)}}\cos
(\int_{0}^{z}q(z^{\prime })dz^{\prime })  \nonumber \\
\varphi _{o}^{-n}(z) &\approx &\frac{1}{\sqrt{q(z)}}\sin
(\int_{0}^{z}q(z^{\prime })dz^{\prime })\text{,}  \label{WKB}
\end{eqnarray}
where $q(z)=\sqrt{k^{2}+\kappa _{n}^{2}u(z)}$. It is well known that Eqs. 
\ref{WKB} are valid when the condition

\begin{equation}
\left| \frac{dq(z)}{dz}\right| \ll q(z)^{2}  \label{WKBcond}
\end{equation}
is fulfilled\cite{Morse}. This condition may be satisfied for the sinusoidal
mode and for any $z$, whenever\label{WKBpar} 
\begin{equation}
\xi \equiv \frac{\pi ^{2}(\kappa _{n}/k)^{3}}{4\kappa _{n}L}\ll 1.
\label{csi}
\end{equation}

In this case the semiclassical solutions (Eqs. \ref{WKB}\ ) are valid for
all $z$ and the reflection coefficient $r_{-n}(k)$ vanishes as $\xi $ goes
to zero, as we will show below. When $\xi \gtrsim 1$ the inequality \ref
{WKBcond} is not valid near the regions $|z|\lesssim L/2$. In this case the
modulus of $r_{-n}(k)$ may increase even to $1.$ Notice that we may have $%
\xi >1$ even for large $kL$ and $\kappa _{n}L.$ For example, for $\kappa
_{n}L\sim 10^{5}$ and $kL\sim 10^{3},$ $\xi \approx 24$.

Near $z=-L/2$, $u(z)$ may be approximated by a straight line and the
solutions of Eq. \ref{Schr} may be written as:

\begin{equation}
\varphi _{e,o}^{-n}(z)\propto w^{1/3}(A_{e,o}J_{1/3}(w)+B_{e,o}J_{-1/3}(w))
\label{linear}
\end{equation}
where

\begin{equation}
w=\frac{\pi }{6}\kappa _{n}L\left( \frac{2z}{L}+1+\frac{4k^{2}}{\pi
^{2}\kappa _{n}^{2}}\right) ^{3/2}.
\end{equation}

The logarithmic derivatives at $z=-L/2$ are given by:

\begin{equation}
\beta _{e,o}^{-n}=k\left( \xi +\frac{A_{e,o}J_{1/3}^{\prime }(\xi
^{-1}/3)+B_{e,o}J_{-1/3}^{\prime }(\xi ^{-1}/3)}{A_{e,o}J_{1/3}(\xi
^{-1}/3)+B_{e,o}J_{-1/3}(\xi ^{-1}/3)}\right) .  \nonumber
\end{equation}
By connecting the even and odd solutions of Eq. \ref{linear} with those
given by Eqs. \ref{WKB} , we obtain :

\begin{equation}
\frac{A_{e}}{B_{e}}=\chi (\varphi )=-\frac{\text{sin}(\varphi +\pi /12)}{%
\text{cos}(\varphi -\pi /12)}
\end{equation}
and $A_{o}/B_{o}=\chi (\varphi +\pi /2),$ where

\begin{equation}
\varphi \simeq \int_{0}^{-L/2}\sqrt{k^{2}+\kappa _{n}^{2}u(z)}dz
\end{equation}

For $\xi \ \ll 1$, we use the asymptotic expressions for $J_{\pm 1/3}(\xi
^{-1}/3)$ to obtain $\beta _{e}^{-n}\approx -k(\xi /2+\tan \varphi )$ and $%
\beta _{o}^{-n}\approx -k(\xi /2-\cot \varphi ).$ From Eqs. \ref{reflec} and 
\ref{trans1} we obtain the same result (up to a phase coming from our
definition of $r_{-n}$ and $t_{-n}$) of ref. \cite{scully2}:

\begin{eqnarray}
r_{-n} &=&-i(\xi /4)\left[ \cos 2\varphi +(\xi /4)\sin 2\varphi \right]
t_{-n}  \nonumber \\
&& \\
t_{-n} &=&e^{-ikL}\left[ (\xi /4)^{2}e^{-2i\varphi }+(1+i\xi
/4)^{2}e^{2i\varphi }\right] ^{-1} .  \nonumber
\end{eqnarray}

For $\xi \gg 1$, we may use the series expansion for $J_{\pm 1/3}(\xi
^{-1}/3)$ to obtain $\beta _{e}^{-n}=-\alpha k\xi ^{1/3}\chi (\varphi )$ and 
$\beta _{o}^{-n}=\alpha k\xi ^{1/3}\chi (\varphi +\pi /2)$ where $\alpha =$ $%
(2/9)^{1/3}\Gamma (2/3)/\Gamma (4/3).$ In this case we get: 
\begin{equation}
r_{-n}=\frac{-i\alpha \xi ^{1/3}(\chi (\varphi )+\chi (\varphi +\pi /2))}{
(1+i\alpha \xi ^{1/3}\chi (\varphi ))(1-i\alpha \xi ^{1/3}\chi (\varphi +\pi
/2))}e^{-ikL}
\end{equation}
and 
\begin{equation}
t_{-n}=\frac{1-\alpha ^{2}\xi ^{2/3}\chi (\varphi )\chi (\varphi +\pi /2)}{
(1+i\alpha \xi ^{1/3}\chi (\varphi ))(1-i\alpha \xi ^{1/3}\chi (\varphi +\pi
/2))}e^{-ikL}.
\end{equation}

Resonances occur when $\beta _{e}^{-n}$, or $\beta _{o}^{-n}$, are zero,
that is when either $\chi (\varphi )$ or $\chi (\varphi +\pi /2)$ are zero,
which give us a simple condition for localizing them:

\begin{equation}
\frac{\kappa _{n}L}{\pi }\int_{0}^{\pi /2}\sqrt{(k/\kappa _{n})^{2}+(\pi /2) 
\text{cos}(\theta )}d\theta =m\pi /2+\pi /12\text{,}
\end{equation}
with $m=0,1,2,...$ For $(k/\kappa _{n})^{2}\ \ll 1$ we may evaluate the
integral approximately and get for the resonance positions:

\begin{equation}
\kappa _{n}L\sim 2.092\times (m\pi /2+\pi /12).
\end{equation}

Therefore, for the sinusoidal field mode we will also obtain a series of
resonances at values of $\kappa _{n}L$ that are separated by an interval
approximately equal to $\pi $. In addition their positions are shifted in
relation to the constant field case, their widths are wider and increase
with $\kappa _{n}L$. We will show that these results give rise to new
features in the physical quantities of interest.

\section{Numerical results}

We have checked that the WKB-like eigenfunctions agrees extremely well with
the numerical solutions at each point, even for low values of $\kappa _{n}L$
and $k/\kappa _{n}$. For higher values of $\kappa _{n}L$ (realistic values
included) the WKB-like\ solution should be increasingly better.

For very small values of $\kappa _{n}L$ the resonances in $P_{\text{em}}^{n}$
, although wider than in the constant field case, do not correspond to
appreciable values of $P_{\text{em}}^{m}$ for $m\neq n$. For large values of 
$\kappa _{n}L$ the resonances in $P_{\text{em}}^{n}$ are wide enough to
produce appreciable values of $P_{\text{em}}^{m}$ for $m\neq n$ , so one
should take them in account when predicting the total probabilities $P_{%
\text{em }}$, or $T$. This is not the case for the constant field mode where
the narrow resonances assure a similar behavior of $P_{\text{em}}^{n}$ and $%
P_{\text{em}}$, $T^{n}$ and $T$, except for accidental numerical
coincidences.

In Fig. 1 we show our results for the probability of emission of a photon
when there is zero photons inside the cavity (vacuum field), $P_{\text{em}
}^{o}$, for a fixed value of $k/\kappa _{n}=0.01,$ as a function of $\kappa
_{n}L$ in the range $100\pi <\kappa _{n}L<104\pi $. We see that the
resonances are shifted and wider in the sinusoidal mode (solid line), in
comparison with the constant mode (dashed line), and that their widths are
still much smaller than the resonance separation.

We now consider that the cavity is in contact with a heat bath at a
temperature associated with a thermal mean photon number $n_{b}$. Assuming,
as usual, that we may treat gain and loss independently and that the
interval of the atom separation obeys a Poissonian statistics, it can be
shown that the stationary photon distribution is given by\cite{scully}

\begin{eqnarray}
&&\frac{dp_{n}}{dt}=r(p_{n-1}P_{\text{em}}^{n-1}-p_{n}P_{\text{em}}^{n})- 
\frac{\omega }{Q}(n_{b}+1)  \nonumber \\
&&[np_{n}-(n+1)p_{n+1}]-\frac{\omega }{Q}n_{b}[(n+1)p_{n}-np_{n+1}]\text{.}
\label{master}
\end{eqnarray}
Here $r$ is the atomic injection rate, $Q$ is the cavity quality factor, $%
\omega $ is the cavity resonance frequency and $n_{b}$ is the mean number of
photons inside a cavity in thermal equilibrium with a reservoir at
temperature $T_{b}$. From Eq. \ref{master} we obtain the stationary
population :

\begin{equation}
p_{n}=p_{0}\prod \frac{n_{b}+N_{ex}P_{\text{em}}^{j-1}/j}{(n_{b}+1)},
\label{population}
\end{equation}
where $N_{ex}=rQ/\omega $. This result is similar to that obtained in the
conventional micromaser, where the value of $P_{\text{em}}^{n}$ is given by
sin$^{2}(\tau \kappa _{n}^{2}/(2m))$, $\tau =(m/\hbar k)L$ being the time of
flight of the atom through the cavity. For $k/\kappa _{0}\ \ll 1$ the photon
distribution obtained from Eq. \ref{population} is completely different from
that of the conventional micromaser. In this case the predictions are
strongly dependent on the field mode profile as long as $n_{b}$ is small
enough and $N_{ex}$ large enough.

Fig. 2 shows the stationary photon distributions, for $N_{ex}=1000$, $n_{b}=1
$ and $k/\kappa _{0}=0.01$, when the $\kappa _{2}L$ value corresponds to the
position of the $100^{th}$ resonance in the case of the constant mode $%
(\kappa _{2}L=99\pi )$ and of the sinusoidal mode $(\kappa _{2}L\approx
103.7\pi )$. Fig 2a (constant mode) shows the apparition of two thermal
distributions with peaks at $n=3,12.$ For such low values of $\kappa L$ this
may happen in two special cases \cite{scully2}, when there is a resonance as
in $\kappa L\sqrt[4]{3}=(99\pi /3^{1/4})(3)^{1/4}=99\pi $ $(n=3)$ or when
there is a numerical coincidence as in $\kappa L\sqrt[4]{12}=(99\pi
/3^{1/4})(12)^{1/4}\approx 140.007\pi $ $(n=12)$. Otherwise we would obtain
an equilibrium thermal distribution at the temperature corresponding to the
average photon number $n_{b}$. This is not the case for the sinusoidal mode
(see Fig. 2b) since the probability of photon emission, $P_{\text{em }}^{j}$%
, may be important for several different values of $j$ besides the resonant
probability for $n=0$, even at these low values of $\kappa L$. This is due
to the large width of the resonances for the sinusoidal mode, in contrast
with the sharp ones for the constant field case. As we mentioned before,
these widths increase with $\kappa L$ for the sinusoidal mode, producing an
extremely complex photon statistics.

Realistic values for the coupling of Rydberg atoms with microwave cavities
are much larger than the values considered above. For example, consider the
Rydberg transition between circular states with principal quantum numbers $%
50 $ and $51$ in Rubidium at $51GHz$, used by the Ecole Normale
Sup\'{e}rieure Group in recent experiments\cite{ENS}. For a cylindrical
cavity, with a length $L\approx 0.40cm$ and a radius $R\approx 0.68cm$, this
transition is resonant with the TE$_{121}$mode. For a dipole moment $%
d\approx 10^{-26}Cm$, we get $\kappa _{0}\approx 2.1\times 10^{5}cm^{-1}$and 
$\kappa _{0}L\approx 27000\pi $.

In Fig. 3 we plot the probability of detecting one transmitted atom, $%
T^{n}=T_{e}^{n}+T_{f}^{n}$, and the probability that one photon be emitted, $%
P_{\text{em}}^{n}$$=R_{f}^{n}+T_{f}^{n}$, if we have initially $n=0,1,2,3$
in the cavity, as a function of $\kappa _{0}L$, for $30000\pi <\kappa
_{0}L<30005\pi $ and $k/\kappa _{0}=0.01$. This ratio corresponds to
Rubidium atom velocities of approximately $0.2mm/s$. Atomic velocities lower
than that have been already obtained by evaporative cooling \cite
{evaporative}. For these velocity values the injection rate should be very
small, if we wish to have only one atom at a time in the cavity ($r\simeq
0.06s^{-1})$, and the value of $N_{ex}$ would be smaller than $1$ even for $%
Q\approx 3$$\times 10^{12}!$ In this case the stationary photon distribution
would be very close to a thermal one, inhibiting the new features shown in
Fig. 2. In Fig. 3, after a careful inspection, we note that $T^{n}$ and $P_{%
\text{em}}^{n}$ show a similar, but not identical, behavior. $P_{\text{em}
}^{n}$ presents better resolved resonances than $T^{n}$ and this could be
useful for determining the more convenient experimental setup.

In Fig. 4 we plot the total probability of detecting the transmitted atom, $%
T $, for the sinusoidal mode (dotted line) and the constant mode (solid
line), when a coherent state, either with $\overline{n}=0.25$ or $\overline{%
n }=2$, is initially present inside the cavity. The figure shows very sharp
resonances in $\kappa _{0}L$ for the constant mode. They are a consequence
of the fact that only one resonance is contributing for a given $n$, with a
statistical weight $|c(n)|^{2}$. For the sinusoidal mode the resonances in $%
T^{n}$ are broader and several resonances may contribute for a fixed $\kappa
_{0}L.$ The case in which $\overline{n}=2$ is much more sensible to these
features, showing, for the constant mode, many little resonance peaks and,
for the sinusoidal mode, very bad resolved resonances.

\section{Conclusions}

The behavior of ultracold atoms interacting with a field inside a high
quality microwave cavity depends in an essential way on the profile of the
mode inside the cavity. For warm atoms it is the value of the Rabi angle and
the rate of incident atoms that defines the gain. For ultracold atoms we
should also consider the exchange of momentum between the cavity and the
atom. Most atoms are reflected when the atom-field system is in the state $%
|+n\rangle $, independently of the shape of the mode. When the atom-field
system is in the state $|-n\rangle $ and the shape of the cavity is smooth
enough, so that the effective de \ Broglie wavelength does not vary
appreciably through the cavity, most atoms would be transmitted. When this
is not the case the atoms would be reflected, unless $\kappa _{n}L$ is close
to certain resonant values.

We have calculated, in the case of a sinusoidal mode, the probability of
transmission of an ultracold atom and the probability of emission of a
photon comparing them with the corresponding mesa function results. In the
case of a mesa function sharp resonances appear in the transmission
coefficient as a function of $\kappa _{n}L$. In the case of the sinusoidal
mode resonances are present when the value of $\frac{\pi ^{2}(\kappa
_{n}/k)^{3}}{4\kappa _{n}L}\gtrsim 1,$ even when $\kappa _{n}L$ is very
large. This may give us a chance to test these effects in a realistic
experiment (when $\kappa _{0}L\sim 10^{5})$ using a cylindrical cavity. We
also found, in the case of the sinusoidal mode, that the width of the
resonances are larger, but still well resolved for realistic parameters, and
increases with $\kappa _{0}L$, as long as $k/\kappa _{0}$ is very small. As
a consequence, the transmission of atoms through a cavity containing a
coherent state with few photons, as a function of $\kappa _{0}L$, has a
completely different shape for the two modes we have studied. The broadening
of the resonances produces also a more complex photon statistics for the
stationary state, at least for theoretical parameters, to our knowledge, far
from realistic ones.

These effects show the importance of considering the details of the cavity
mode when analyzing the scattering of ultracold atoms by a resonant cavity.

We should remark that a serious difficulty in realizing these experiments is
the long time of flight that ultracold atoms need to reach the cavity after
their production.

\section{Acknowledgments}

This work is supported in part by Centro Latino Americano de F\'{\i }sica
(CLAF), The Brazilian Conselho Nacional de Desenvolvimento Cient\'{\i }fico
e Tecnol\'{o}gico (CNPq), Programa de Apoio a N\'ucleos de Excel\^encia
(PRONEX) and Funda\c{c}\~{a}o Universit\'{a}ria Jos\'{e} Bonif\'{a}cio
(FUJB).

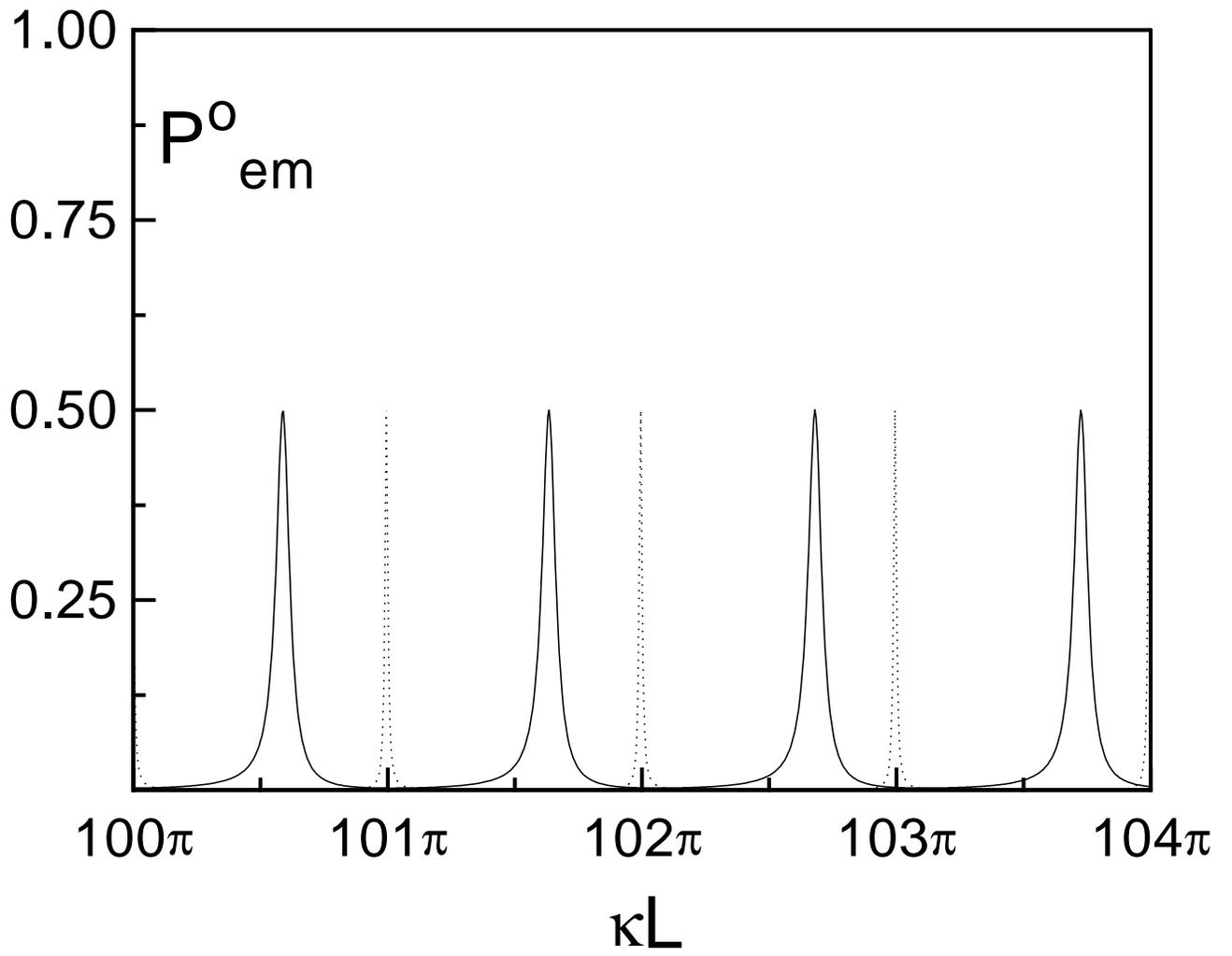
\begin{figure}[tbp]
\caption{Probability of emission of a photon when the cavity is initially
empty as a function of $\kappa L$ when the mode is: the sinusoidal mode
(solid line) and the constant mode (dotted line)}
\end{figure}

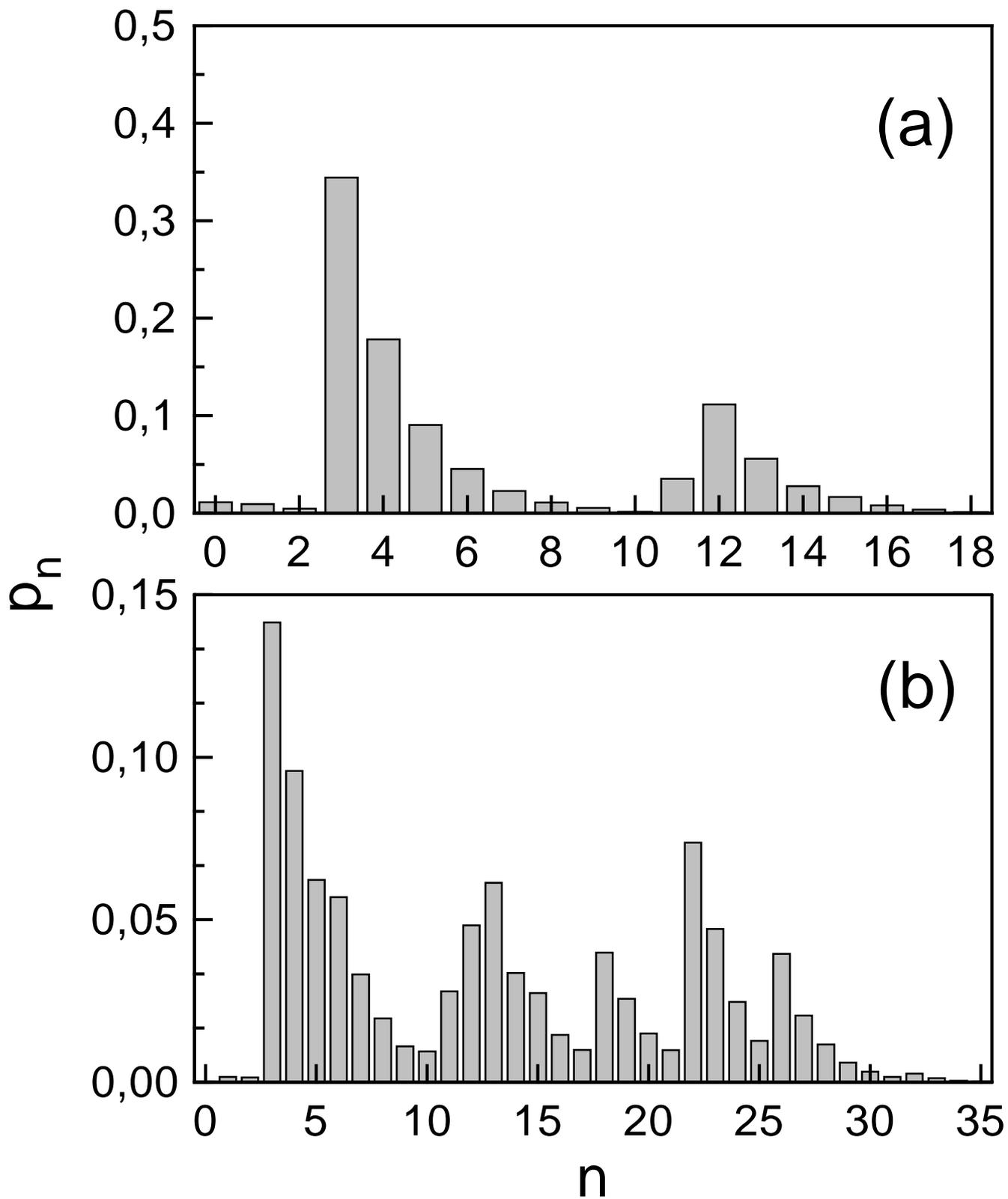
\begin{figure}[tbp]
\caption{Stationary photon distribution for $N_{ex}=1000$, $k/\kappa =0.01$
and $n_b=1$ when $\kappa_2 L $ corresponds to the $100^{th}$
resonance:(a)constant mode; (b) sinusoidal mode.}
\end{figure}

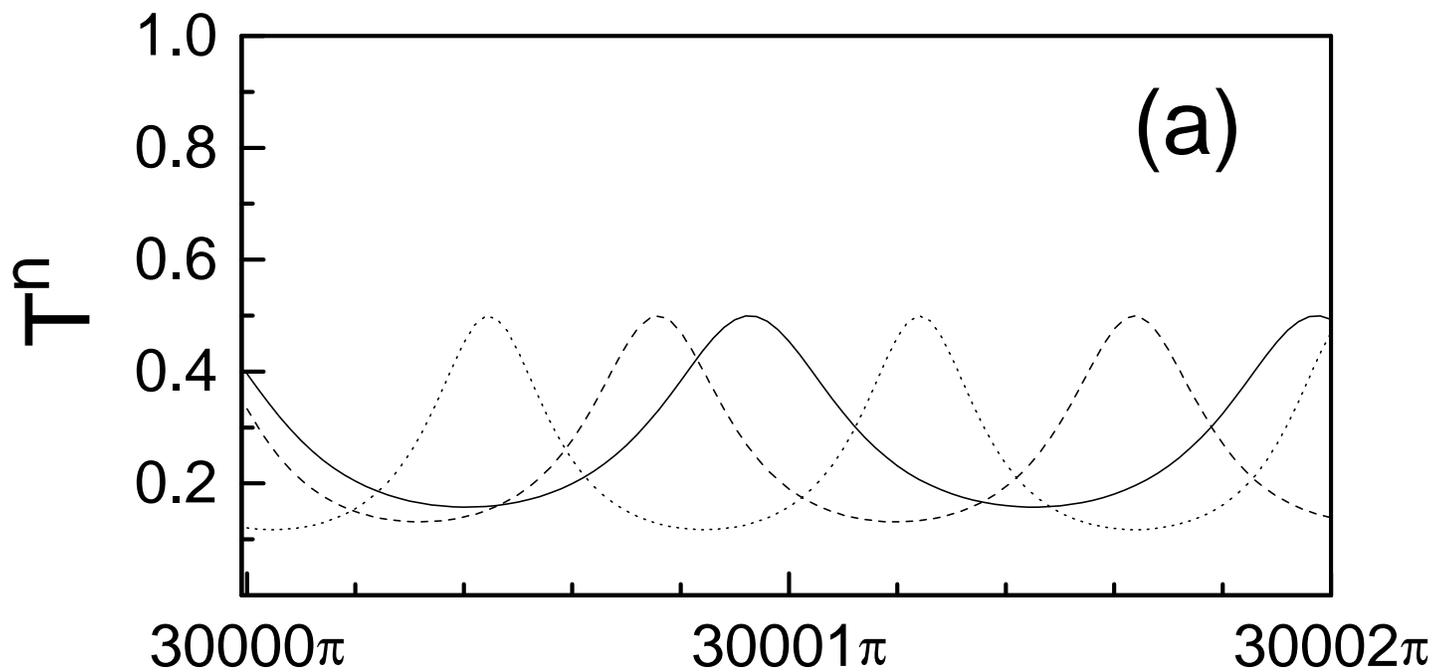
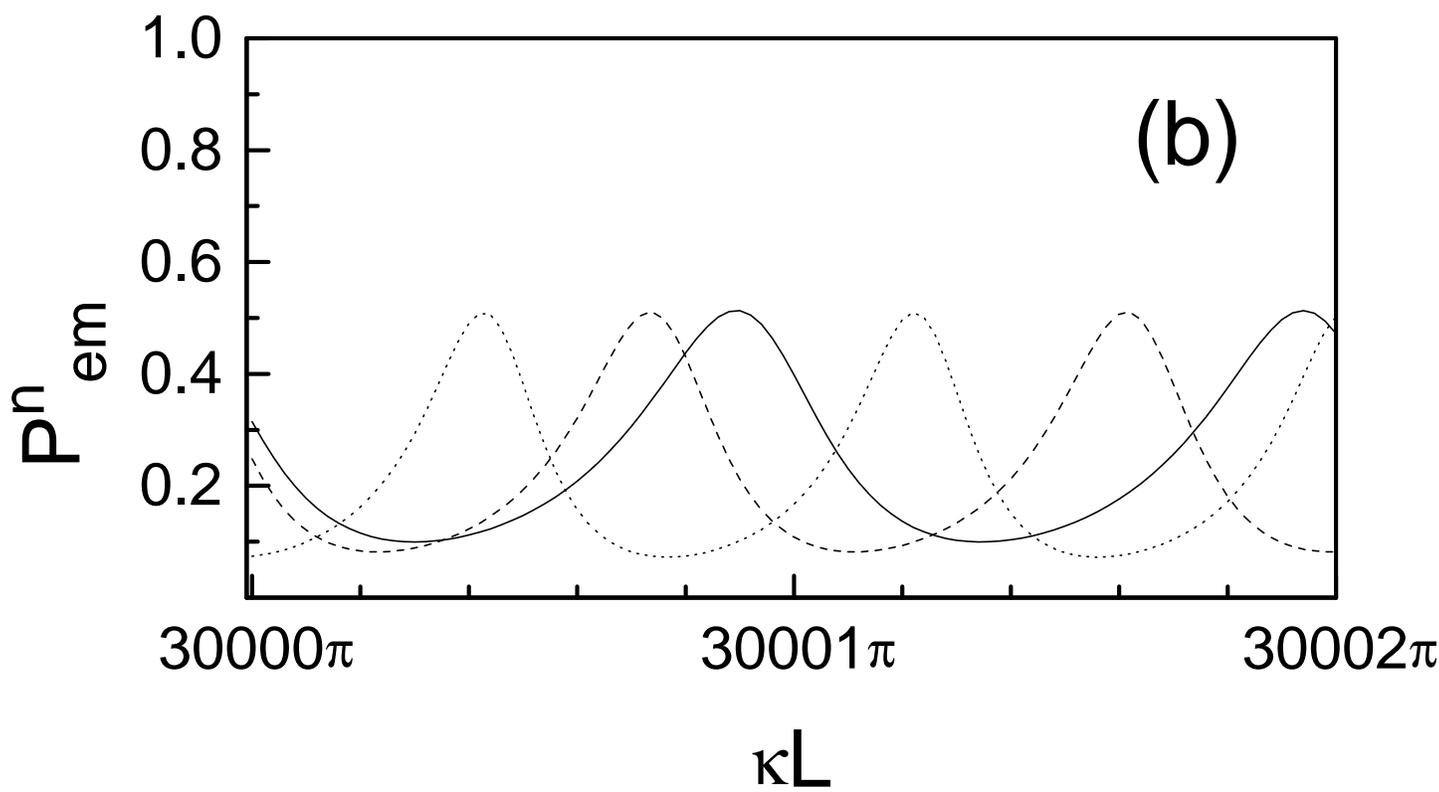
\begin{figure}[tbp]
\caption{(a)Probability that an atom being transmitted through the cavity
and (b)Probability of emission of a photon when the cavity has initially $n$
photons, as a function of $\kappa L$: $n=0$ (solid line), $n=1$ (dashed
line), $n=2$ (dotted line). $k/\kappa =0.01$. Sinusoidal mode}
\end{figure}

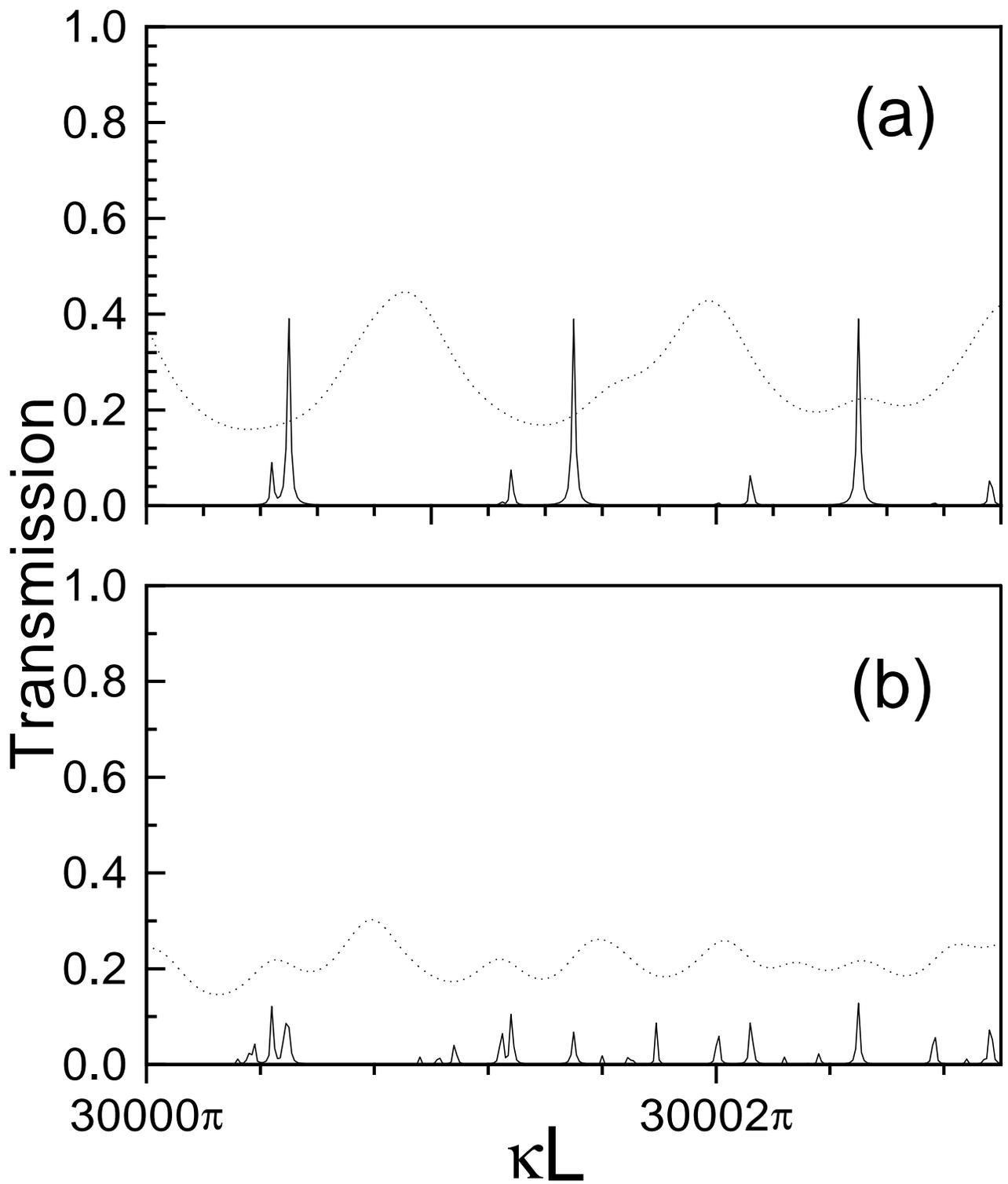
\begin{figure}[tbp]
\caption{Probability that an atom being transmitted when initially there is
a coherent state in the cavity as a function of $\kappa L$ for the constant
mode (full line) and the cosine mode (dashed line ) (a) : $\bar n=0.25$
(solid line); (b): $\bar n=2$. $k/\kappa =0.01$. Sinusoidal mode}
\end{figure}

\end{document}